\documentclass[12pt]{iopart}
\usepackage{iopams}
\usepackage{epsfig}
\usepackage{graphics}

\begin{document}
\title{Robustness and epistasis in mutation-selection models}
\author{Andrea Wolff and Joachim Krug}
\address{Institut f\"ur Theoretische Physik, Universit\"at zu K\"oln, Z\"ulpicher Str. 77, 50937 K\"oln, Germany}
\eads{\mailto{awolff@thp.uni-koeln.de} and \mailto{krug@thp.uni-koeln.de}}

\begin{abstract}
We investigate the fitness advantage associated with the robustness of a phenotype against deleterious mutations using deterministic mutation-selection models of quasispecies type equipped with a mesa shaped fitness landscape. We obtain analytic results for the robustness effect which become exact in the limit of infinite sequence length. Thereby, we are able to clarify a seeming contradiction between recent rigorous work and an earlier heuristic treatment based on a mapping to a 
Schr\"odinger equation. We exploit the quantum mechanical analogy to calculate a correction term for finite sequence lengths and verify our analytic results by numerical studies. In addition, we investigate the occurrence of
an error threshold for a general class of epistatic landscape and show that diminishing
epistasis is a necessary but not sufficient condition for error threshold behavior.
\end{abstract}
\section{Introduction}
In current evolutionary theory, the concept of \textit{robustness},
referring to the invariance of the phenotype under 
pertubations, is of central importance \cite{deVisser2003,Krakauer2002}.
Here we address specifically \textit{mutational robustness}, which we take to imply
the stability of some biological function with respect to mutations away from the 
optimal genotype. To be precise, suppose the genotype is encoded by a sequence of length
$L$, and the number of mismatches with respect to the optimal genotype is denoted by $k$.
Robustness is then quantified by the maximum number of mismatches $k_0$, that can be tolerated
before the fitness of the individual falls significantly below that of the optimal genotype
at $k=0$. This situation arises e.g. in the evolution of regulatory motifs, where the fitness 
is a function of the binding affinity to the regulatory protein \cite{Gerland2002,Berg2004}.
Assuming that the fitness is independent of $k$ both for $k \leq k_0$ and for $k > k_0$,
the fitness landscape has the shape of a \textit{mesa} parametrized by its width
$k_0$ and height $w_0$ \cite{Peliti2002}. 

We consider deterministic mutation-selection models of quasispecies type, which describe the dynamics
of large (effectively infinite) populations \cite{JK2006}. We analyse the stationary state of 
mutation-selection balance, focusing on the dependence of the population fitness on the parameters
$k_0$ and $w_0$. This allows us to identify the conditions under which a broad fitness peak of relatively
low selective advantage outcompetes a higher but narrower peak \cite{Schuster1988}, 
a phenomenon that has been referred to as the \textit{survival of the flattest} 
\cite{Wilke2001,Codoner2006,Sanjuan2007,Rolland2007,Sardanyes2008}. Our central aim is to obtain
analytic results for the robustness effect that become exact in the limit of long
sequences. In particular, we want to clarify whether the selective advantage is a
function primarily of the \textit{relative} number of tolerable mismatches $x_0 = k_0/L$,
or of the total number of mismatches $k_0$.

Robustness in the sense described above is a special case of \textit{epistasis}, which refers
more generally to any nonlinear relationship between the number of mutations away from the 
optimal genotype and the corresponding fitness effect \cite{Phillips2000}. A simple way to parametrize epistasis
is to let the loss of fitness vary with the number of mismatches as $k^\alpha$, such that the non-epistatic case
$\alpha = 1$ separates regimes of synergistic ($\alpha > 1$) and diminishing 
($\alpha < 1$) epistasis \cite{Wiehe1997,Jain2008}. An important problem in previous work 
on mutation-selection models has been to identify the conditions under which epistatic
fitness landscapes display an \textit{error threshold},  a term that refers to the discontinuous
delocalization of the population from the vicinity of the fitness peak as the mutation
rate is increased beyond a critical value \cite{JK2006,Hermisson2002}. Improving on earlier work that found that
only landscapes with diminishing epistasis ($\alpha < 1$) have an error threshold, we derive
here the more stringent condition $\alpha \leq 1/2$ on the epistasis exponent.

\subsection{Organization of the article}

We base our work on two complementary analytic approaches. First, 
recent progress in the theory of mutation-selection models 
\cite{Peliti2002,Hermisson2002,Baake2005,Saakian2006,Park2006,Saakian2007,Baake2007,Sato2007}
provides an expression for the population fitness in terms of a
maximum principle (MP) that becomes exact when the
limit $L \to \infty$ is performed keeping the ratio
$x_0 = k_0/L$ fixed. Second, Gerland and Hwa (GH) \cite{Gerland2002}
have used a drift-diffusion approximation to map the
mutation-selection problem onto a one-dimensional Schr\"odinger
equation which is then analyzed with standard techniques. 

Our work was initially motivated by the observation of a discrepancy
between the two approaches: Whereas the MP  predicts that the
selective advantage of a broad mesa should vanish when the limit $L
\to \infty$ is taken at fixed $k_0$, in the GH approach a finite
selective advantage is retained in this limit, which 
depends on the absolute value of $k_0$ rather than on $x_0$. 
After introducing the model and briefly reviewing the results of the 
MP approach in section \ref{Models}, we therefore provide a detailed
discussion of the drift-diffusion approximation used by GH in section \ref{Continuum}.
We emphasize that it amounts to a harmonic approximation, 
and show how it can be improved in such a way
that the results based on the MP are recovered. 

The mapping
to one-dimensional quantum mechanics is nevertheless useful, as it 
allows us to derive the leading
finite size correction to the population fitness. As a consequence we find
excellent agreement between the analytic predictions and numerical solutions 
of the discrete mutation-selection equations. In section \ref{Selection} we
consider the selection transition in a two-peak landscape first studied by 
Schuster and Swetina \cite{Schuster1988,Tarazona1992}, in which the population shifts from a high,
narrow fitness maximum to a lower but broader peak with increasing
mutation rate. 
The occurrence of this transition is an indicator for the superiority
of robustness over fitness in certain parameter regimes.
In section \ref{epistasis} we use the MP approach to derive the critical
value of the epistasis exponent $\alpha$ and verify our prediction by
numerical calculations. 
Finally, some conclusions are presented in section \ref{Conclusions}. 
Details of the derivation of the improved continuum approximation and the
generalization to arbitrary alphabet size can be found in two
appendices.
  
\section{Mutation-selection models and the maximum principle}
\label{Models}

We consider the simplest case of binary sequences and adopt continuous time dynamics
of the Crow-Kimura type, in which the mutation and selection terms act in parallel
\cite{JK2006}. Point mutations occur at rate $\mu$, and the (Malthusian) fitness
is assumed from the outset to be a function $w_k$ only of the Hamming distance
$k$ to the optimal sequence at $k=0$. The population structure is described by 
the fraction $P_k(t)$ of individuals with $k$ mismatches, which satisfies the evolution
equation
\begin{equation}
\label{CK}
\frac{dP_k}{dt} = (w_k-\bar w)P_k + \mu(k+1) P_{k+1} + \mu (L - k + 1) P_{k-1}
- \mu L P_k.
\end{equation}
with $1 \leq k \leq L-1$ and obvious modifications for $k=0$ and $k=L$. The nonlinearity
introduced by the mean fitness $\bar w(t) = \sum_k w_k P_k$ can be eliminated by passing
to unnormalized population variables \cite{JK2006,Thompson1974}. 
At long times the population distribution 
therefore approaches the principal eigenvector $P^\ast_k$
of the linear dynamics, which is the solution
of the eigenvalue problem
\begin{equation}
\label{Eigen}
\Lambda P^\ast_k = (w_k - \mu L) P^\ast_k + \mu(k+1) P^\ast_{k+1} + 
\mu (L - k + 1) P^\ast_{k-1}
\end{equation}
with the maximal eigenvalue $\Lambda$. This eigenvalue 
is equal to the long-time limit of the mean population fitness $\bar w$,
and it is the main quantity of interest in this paper. Depending on
the context we will refer to $\Lambda$ as the mean population fitness,
the population growth rate, the principal eigenvalue of the
mutation-selection matrix defined by (\ref{Eigen}) 
or the ground state energy of the corresponding quantum mechanical
problem, to be defined in subsection \ref{QuantumMapping}.

A considerable body of work has been devoted to the solution of (\ref{Eigen}) for 
large $L$. In order to obtain nontrivial behavior in the limit $L \to \infty$,
it is necessary to either scale the mutation rate $\sim 1/L$ or the fitness 
$\sim L$. We adopt here the first choice and take $L \to \infty$, 
$\mu \to 0$ with $\gamma = \mu L$ fixed. If, in addition, the fitness landscape
$w_k$ is assumed to depend only on the relative number of mismatches, such that
\begin{equation}
\label{fit}
w_k = f(x), \;\;\; x = k/L
\end{equation}
the principal eigenvalue in (\ref{Eigen}) is given, for $L \to \infty$,
by the solution of 
a one-dimensional variational problem as 
\cite{Hermisson2002,Baake2005,Saakian2006,Park2006,Saakian2007,Baake2007}
\begin{equation}
\label{maximum}
\Lambda = \max_{x \in [0,1]} \{f(x) - \gamma [1 - 2 \sqrt{x(1 - x)}]\};
\end{equation}
see subsection \ref{secimprovement} and Appendix A for a heuristic derivation, 
and Appendix B for the generalization to arbitrary alphabet size.
Moreover, if $f(x)$ is differentiable the leading order correction to (\ref{maximum}) takes
the form \cite{Park2006,Saakian2007}
\begin{equation}
\label{corr}
\Delta \Lambda = \frac{\gamma}{2 L \sqrt{x_c - x_c^2}} [1 -
\sqrt{1 - 2 f''(x^\ast) (x_c -x_c^2)^{3/2}/\gamma}],
\end{equation}
where $x_c$ is the value at which the maximum in (\ref{maximum}) is attained.

In the first part of this paper we focus on mesa landscapes of the form
\begin{equation}
\label{mesa}
w_k = \left\{ \begin{array}{l@{\quad:\quad}l}
w_0 > 0  & 0 \leq k \leq k_0 \\ 
0  &  k > k_0, \end{array} \right.
\end{equation}
where $w_0$ is the selective advantage of the functional phenotype and $k_0$ denotes the 
number of tolerable mismatches. 
Within the class of \textit{scaling} landscapes (\ref{fit}), this is realized
by setting 
\begin{equation}
\label{theta}
f(x) = w_0 \theta(x - x_0), 
\end{equation}
where $\theta$ is the Heaviside
step function and $x_0 = k_0/L$. Provided $x_0 < 1/2$, application of
the maximum principle (\ref{maximum}) yields
\begin{equation}
\label{mesa2}
\Lambda = \left\{ \begin{array}{l@{\quad:\quad}l}
w_0 - \gamma (1 - 2 \sqrt{x_0(1 - x_0)})  & w_0 > w_0^c = \gamma (1 - 2 \sqrt{x_0(1 - x_0)})  \\ 
0  &   w_0 < w_0^c. \end{array} \right.
\end{equation}
The value $w_0^c$ of the selective advantage marks the location of the \textit{error threshold} at which
the population delocalizes from the fitness peak and the location $x_c$ of the maximum in (\ref{maximum})
jumps from $x_c = x_0$ to $x_c = 1/2$. 

The expression (\ref{corr}) is clearly not applicable to the
discontinuous mesa landscape (\ref{theta}). In fact we will show below that the leading order correction 
$\Delta \Lambda$ is of order $L^{-2/3}$ or $L^{-1/2}$ rather than
$L^{-1}$ in this case.

\section{Continuum limit and the drift-diffusion equation}
\label{Continuum}

\subsection{Derivation and status}
\label{Status}

A natural approach to analyzing (\ref{CK}) and (\ref{Eigen}) for large $L$ is to perform a continuum
limit in the index $k$. To this end we introduce $\epsilon = 1/L$ as small parameter and
replace the population variable $P_k$ by a function 
\begin{equation}
\label{cont}
\phi(x) = \lim_{L \to \infty} P_{xL}.
\end{equation}
The fitness is taken to be of the general form (\ref{fit}).  
Expanding the finite differences in (\ref{Eigen}) to second order in $\epsilon$ then yields
the stationary drift-diffusion equation
\begin{equation}
\label{drift}
f \phi - \epsilon \gamma \frac{d}{dx} (1 - 2x) \phi + \frac{\epsilon^2 \gamma}{2} \frac{d^2}{dx^2} \phi = \Lambda \phi.
\end{equation}
This is identical to the equation obtained by GH \cite{Gerland2002}, who however write it in terms
of the unscaled variable $k = Lx$. 

Before proceeding with the analysis of (\ref{drift}), some remarks concerning the accuracy 
of the second order expansion are appropriate. In the absence of selection ($f = 0$) the principal
eigenvalue in (\ref{drift}) is readily seen to be $\Lambda = 0$, and the corresponding 
(right) eigenfunction is a Gaussian centered at $x=1/2$,
\begin{equation}
\label{Gauss}
\phi_0(x) \sim \exp[-(1 - 2x)^2/2 \epsilon].
\end{equation}
This is just the central limit approximation to the binomial distribution
\begin{equation}
\label{binomial}
P_k^0 = 2^{-L} {L \choose k}
\end{equation}
which solves (\ref{Eigen}) for $w_k = 0$ and $\Lambda = 0$. It is well known
that the central limit approximation of (\ref{binomial}) 
is accurate in a region of size
$\sqrt{L}$ around $k = L/2$, but becomes imprecise for deviations of order
$L$. An improved approximation is provided by the theory of large deviations
\cite{Sornette2000}, in which the ansatz
\begin{equation}
\label{large}
P_k \sim \exp[-L u(x)]
\end{equation}
is made to obtain an expression for the large deviation function $u(x)$. In
the context of mutation-selection models, this approach has recently been
introduced by Saakian \cite{Saakian2007}, who showed that it allows to 
derive the exact relation (\ref{maximum}) in a relatively
straightforward manner (see Appendix A). 
Equivalent results can be obtained by continuing
the expansion in (\ref{drift}) to all orders in $\epsilon$ and treating
the resulting equation in a WKB-type approximation, which essentially 
corresponds to the ansatz (\ref{large}), see \cite{Sato2007}.  

We conclude that the drift-diffusion approximation (\ref{drift})
can be expected to be quantitatively accurate only near the center $x = 1/2$
of the sequence space. We will nevertheless adhere to this approximation in 
following three subsections, because it allows us to make contact with the work of GH
and to formulate the eigenvalue problem (\ref{Eigen}) 
in the familiar language of one-dimensional quantum mechanics. In subsection \ref{secimprovement} we then show how to go beyond the second order approximation.

\subsection{Mapping to a one-dimensional quantum problem}
\label{QuantumMapping}
The key step in reducing (\ref{drift}) to standard form
is to symmetrize the linear operator on the left hand side, thus
eliminating the first-order drift term. This can be achieved
by the transformation
\begin{equation}
\label{psi}
\phi(x) = \sqrt{\phi_0(x)} \, \psi(x),
\end{equation}
with $\phi_0(x)$ from (\ref{Gauss}), which leads to the stationary Schr\"odinger equation 
\begin{equation}
\label{QM}
- \frac{\epsilon^2 \gamma}{2} \frac{d^2}{dx^2} \psi + V(x) \psi = -(\Lambda - \epsilon \gamma) \psi
\end{equation}
with the effective potential
\begin{equation}
\label{V}
V(x) = \frac{\gamma}{2} (1 - 2 x)^2 - f(x).
\end{equation}
The latter consists of the superposition of a harmonic oscillator centered around $x = 1/2$ with the (negative)
fitness landscape. As pointed out in \cite{Peliti2002}, the inverse sequence length $\epsilon$
plays the role of Planck's constant $\hbar$, 
which implies that the case of interest is the
\textit{semiclassical} limit of the quantum mechanical problem. In particular, for $\epsilon \to 0$
the ground state energy $-\Lambda$ becomes equal to the minimum of the effective potential. 
We thus arrive at the variational principle
\begin{equation}
\label{var}
\Lambda = \max_{x \in [0,1]} [f(x) - \frac{\gamma}{2} (1 - 2x)^2],
\end{equation}
which is precisely the \textit{harmonic approximation} (in the sense of a quadratic expansion around $x=1/2$) of the
exact relation (\ref{maximum}). In this perspective the error threshold corresponds to a shift between
different local minima of $V(x)$, which become degenerate at the transition point. The transition 
is generally of first order, in the sense that the location $x_c$ of the global minimum jumps
discontinuously. Within the harmonic approximation the transition for
the mesa landscape occurs at
\begin{equation}
\label{w0charm}
w_0^c = \frac{\gamma}{2}(1 - 2 x_0)^2 \approx \frac{\gamma}{2} \left(1 - \frac{4 k_0}{L}\right)
\end{equation}
when $x_0 = k_0/L \ll 1$.

\subsection{Semiclassical finite size corrections}
\label{Corrections}

For small but finite $\epsilon$, quantum corrections to the classical limit (\ref{var}) have to be taken
into account. If $f(x)$ is smooth, the ground state wave function is localized near the minimum $x_c$
of the effective potential, and the shift in the ground state energy can be computed by replacing
$V(x)$ by a harmonic well, 
\begin{equation}
\label{harmonic}
V(x) \approx V(x_c) + \frac{1}{2} V''(x_c) (x - x_c)^2 = V(x_c) + \frac{1}{2} [4 \gamma - f''(x_c)]
(x - x_c)^2.
\end{equation}
Identifying $1/\gamma$ with the mass $m$ of the quantum particle [compare to (\ref{QM})], we see that this corresponds to a
harmonic oscillator of frequency $\omega = 2 \gamma \sqrt{1 - f''(x_c)/4\gamma}$. The ground state 
energy $\epsilon \omega/2$, together with the shift $\epsilon \gamma$ on the right hand side of
(\ref{QM}), thus gives rise to the leading order correction
\begin{equation}
\label{QMcorr}
\Delta \Lambda = \frac{\gamma}{L} [1 - \sqrt{1 - f''(x_c)/4 \gamma}],
\end{equation}
which coincides with (\ref{corr}) evaluated for $x_c \approx 1/2$. Similarly, the 
width of the wave function is given by\footnote{Note that, because of the factor
$\sqrt{\phi_0}$ in (\ref{psi}),  this is not equal to the
width of the stationary population distribution.}
\begin{equation}
\label{xi1}
\xi = \sqrt{\gamma \epsilon/2 \omega} = \frac{\sqrt{\epsilon \gamma}}{
[8 \sqrt{1 - f''(x_c)/4 \gamma}]^{1/4}}.
\end{equation}

 In the case of the mesa landscape (\ref{theta}), the potential near $x_c = x_0$ consists
of a linear ramp of slope 
\begin{equation}
\label{slope}
-a = V'(x_0) = 2 \gamma(2 x_0 - 1) < 0
\end{equation}
 followed by a jump of height
$w_0$. For small $\epsilon$, the jump can be considered as effectively infinite (as the kinetic energy of the particle is then very small), and the 
corresponding quantum mechanical ground state problem is standard textbook material
\cite{Rollnik1995}.
>From the solution we obtain the prediction
\begin{equation}
\label{Airy}
\Delta \Lambda = z_1 (\hbar^2/2m)^{1/3} a^{2/3} = 2^{1/3} z_1 \gamma (1 - 2 x_0)^{2/3} L^{-2/3} + {\cal{O}}(L^{-1}),
\end{equation}
where $z_1 \approx -2.33811...$ is the first zero of the Airy
function. The scaling $\Delta \Lambda \sim L^{-2/3}$ was already noted
in \cite{Peliti2002}.    
The width of the wave function can be estimated to be of the order 
\begin{equation}
\label{xiAiry}
\xi \sim (\hbar^2/ma)^{1/3} \sim \epsilon^{2/3}
\end{equation}
in this case.

\subsection{The quantum confinement regime}
\label{Crossover}

We are now prepared to make contact with the approach of GH \cite{Gerland2002}. 
Assuming from the outset that the maximal number of mismatches is small
compared to the sequence length, $1 \ll k_0 \ll L$, they neglect the contribution $2x$ in the 
drift term on the left hand side of (\ref{drift}). The linear operator can then be symmetrized by the
transformation 
\begin{equation}
\label{GHtrafo}
\phi(x) = e^{x/\epsilon} \psi(x), 
\end{equation}
which is obtained from (\ref{psi}) by neglecting the terms quadratic in $x$ in $\phi_0$.  
This leads to a Schr\"odinger problem similar to (\ref{QM}), but with
a potential that differs from $-f(x)$ only by the constant term $\gamma/2$. 
The error threshold is determined by the point at which the decay of the wave function
$\psi(x)$ matches the exponential factor $e^{x/\epsilon}$ in (\ref{GHtrafo}), such that
$\phi(x)$ ceases to be normalizable\footnote{This requires $\psi$ to decay on a scale of order 
unity in unscaled coordinates at the transition, which is actually inconsistent with
the assumption of slow variation on the scale of the sequence index $k$ that underlies the continuum
approximation.}. For $k_0 \gg 1$ the location of the transition is found by GH to be 
\begin{equation}
\label{GHthresh}
w_0^c = \frac{\gamma}{2} \left( 1 + \frac{\pi^2}{k_0^2} \right),
\end{equation}
which depends on the absolute number of mismatches $k_0$, but is independent of $L$.

To reconcile this with the result (\ref{w0charm}), we note that the semiclassical
approximation must break down when the width of the semiclassical wave function,
as estimated in subsection \ref{Corrections}, becomes comparable to the width of the potential
well provided by the fitness function. For the discontinuous mesa landscape this occurs
when 
\begin{equation}
\label{short_mesa}
\xi \sim \epsilon^{2/3} \sim x_0 = \epsilon k_0 \;\; \Rightarrow \;\; k_0 \sim 
\epsilon^{-1/3} = L^{1/3}.
\end{equation}
For a mesa that is shorter than $L^{1/3}$, the energy of the wave function is determined
by its confinement on the scale $x_0$, and it can be estimated from standard
quantum mechanical considerations to be of the order
of $\hbar^2/(m x_0^2) \sim \gamma \epsilon^2/x_0^2 \sim \gamma/k_0^2$. 
For $k_0 \ll L^{1/3}$ this supersedes
the contribution $\sim k_0/L$ on the right hand side of (\ref{w0charm}). We conclude,
therefore, that the leading ``quantum'' correction to the ''classical'' 
eigenvalue $\Lambda = w_0 - \gamma/2$
is a \textit{negative} contribution proportional to $\gamma/k_0^2$, which leads to a 
corresponding positive shift in $w_0^c$, in qualitative 
agreement with (\ref{GHthresh}). For smooth
fitness landscapes the breakdown of the semiclassical regime occurs already at
$k_0 \sim L^{1/2}$, but the condition for the confinement energy contribution 
$\gamma/k_0^2$ to dominate
the $k_0/L$-term in (\ref{w0charm}) still reads $k_0 \ll L^{1/3}$.

\subsection{Beyond the harmonic approximation}
\label{secimprovement}
So far, we have worked in the harmonic approximation around $x=1/2$,
which breaks down near the boundaries $x=0$ and $x=1$. However, to access the
regime $1 \ll k_0 \ll L$ considered by GH, an accurate treatment of the region of small $x\ll 1$
is clearly necessary. In Appendix A we show how the quantum mechanical
treatment can be extended such that it become quantitatively valid
over the whole interval $0 \leq x \leq 1$. 
Based on the considerations of \cite{Saakian2007}, we arrive at the
modified Schr\"odinger equation
\begin{equation}
- \epsilon^2 \gamma \sqrt{x(1-x)} \frac{d^2}{dx^2} \psi +
\left[\gamma(1-2\sqrt{x(1-x)}) - f(x) \right]\psi = - \Lambda \psi,
\label{schroedinger}
\end{equation}
which differs from (\ref{QM}) in two respects. First, the potential
(\ref{V}) is replaced by
\begin{equation}
\label{fullV}
V_\mathrm{full}(x) = \gamma(1-2\sqrt{x(1-x)}) - f(x).
\end{equation}
In the asymptotic limit $\epsilon \to 0$ the principal eigenvalue is
obtained by minimizing $V_\mathrm{full}$, which exactly recovers the
maximum principle (\ref{maximum}). Second, 
the mass of the quantum particle described by (\ref{schroedinger})
becomes position dependent,
\begin{equation}
\label{mass}
m(x) \; \widehat{=} \;  \left(2\gamma\sqrt{x(1-x)}\right)^{-1},
\end{equation}
which replaces the simple identification $m \, \widehat{=} \, 1/\gamma$ in the
harmonic case. Inserting (\ref{fullV}) and (\ref{mass}) 
into the expression (\ref{Airy}) for the finite size correction yields
\begin{equation}
\Delta\Lambda = 2^{-1/3} z_1 \epsilon^{-2/3} \gamma (1-2x_0)^{2/3} [x_0(1-x_0)]^{-1/6}.
\label{eqnDeltaLambdaNew}
\end{equation}
For fixed $x_0$ this still scales as $\epsilon^{2/3} = L^{-2/3}$, but
when taking $L \to \infty$ at fixed $k_0$, such that $x_0 \to 0$, we
find instead that 
\begin{equation}
\label{L1/2}
\Delta\Lambda \to  2^{-1/3} z_1 \gamma x_0^{-1/6} \epsilon^{2/3} = 2^{-1/3}
z_1 \gamma k_0^{-1/6} L^{-1/2}.
\end{equation}

We next revisit the considerations of subsection \ref{Crossover}.   
The width of the ground state wave function is of order 
$\xi \sim (\hbar^2/m a)^{1/3}$, where both $m$ and $a$ now diverge as
$x_0^{-1/2}$ for $x_0 \to 0$. Consequently (\ref{xiAiry}) is replaced by 
\begin{equation}
\label{ximod}
\xi \sim (\epsilon^2 x_0)^{1/3} = \epsilon k_0^{1/3},
\end{equation}
and we see that the condition $\xi \gg \epsilon k_0$ for the breakdown 
of the semiclassical approximation is never be satisfied. We conclude 
that the quantum 
confinement regime discussed in subsection \ref{Crossover} in fact does
not exist, and hence
the improved semiclassical expression (\ref{eqnDeltaLambdaNew}) 
for the finite size correction is expected to remain valid for all
$k_0$ and all $L$, provided that $k_0, L \gg 1$. 

\begin{figure}
\begin{center}
\includegraphics[scale=0.65]{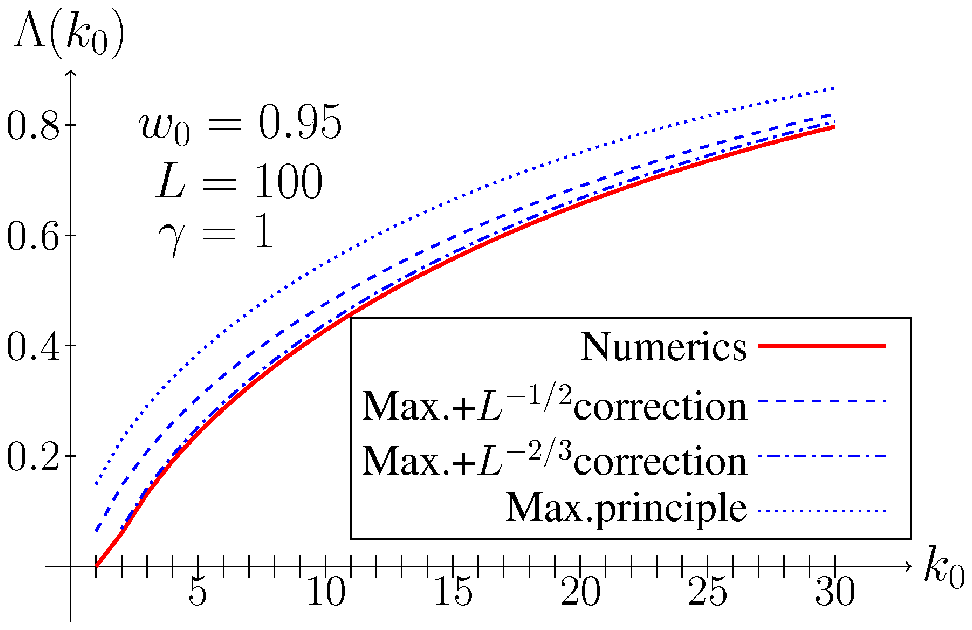}
\includegraphics[scale=0.65]{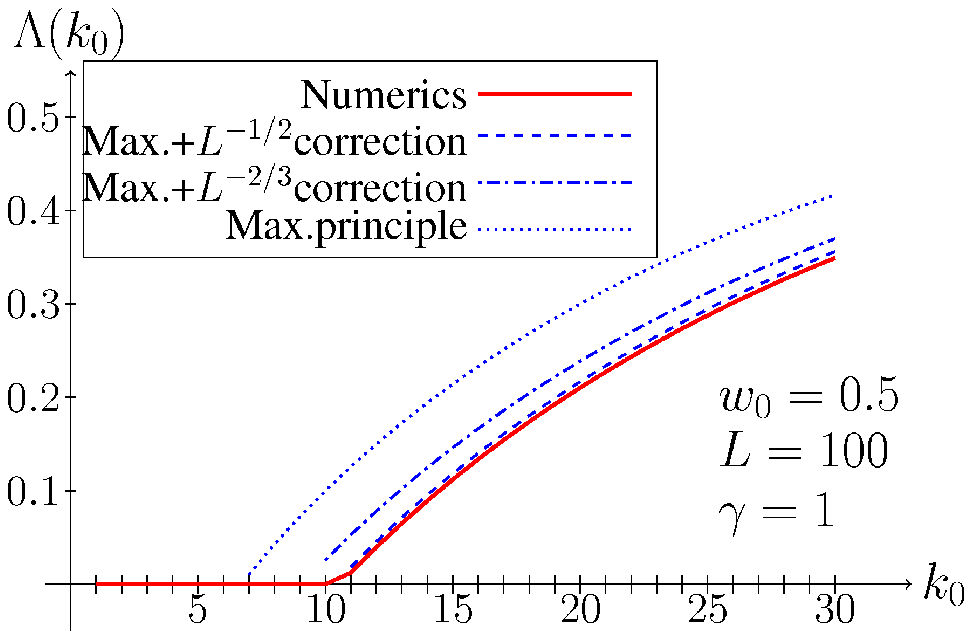}
\caption{Growth rate $\Lambda$ as a function of the plateau width
  $k_0$ for two values of the plateau height $w_0=0.5,0.95$. The
  sequence length is $L=100$ and the mutation rate per sequence is
  $\gamma = 1$. The solution of the maximum principle together with
  the $L^{-1/2}$-correction term (including the position-dependent mass) provides the best agreement with the numerics. The numerical values of the growth rate have been obtained by (numerical) calculation of the largest eigenvalue of the matrix defined by equation (\ref{Eigen}).}
\label{piclambda_von_k}
\end{center}
\end{figure}

\begin{figure}
\begin{center}
\includegraphics[scale=0.65]{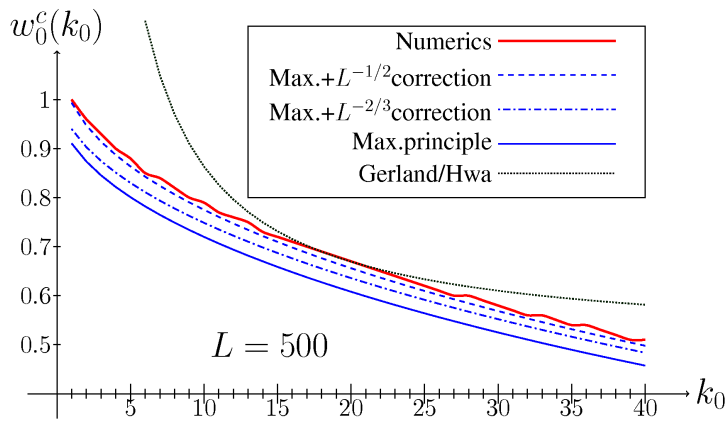}
\caption{Critical plateau height as a function of the plateau width
  $k_0$. The sequence length is $L=500$ the mutation rate per sequence
  $\gamma = 1$. The solution of the maximum principle together with
  the $L^{-1/2}$-correction term provides the best agreement with the
  numerics. With increasing $x_0$, the $L^{-2/3}$ and  
 $L^{-1/2}$-corrections approach each other. The numerical values have
 been obtained by monitoring the average magnetization $M$ defined in
 (\ref{M}) and determining the plateau height, where $M$ jumps from a
 finite value to zero.
The slight modulation of the red line arises from the finite
numerical resolution of this procedure.}
\label{picwoc_von_k}
\end{center}
\end{figure}

\subsection{Numerical results}
\label{Numerics}

To test the analytical predictions derived in the preceding subsections, 
we have carried out a detailed numerical study of the dependence of $\Lambda$ and $w_0^c$
on $k_0$, $L$ and $\gamma$. In figure \ref{piclambda_von_k} we show
two examples for the dependence of $\Lambda$ on the plateau width
$k_0$. The prediction of the asymptotic maximum principle
(\ref{maximum}) reproduces the qualitative behavior of the numerical
data but significantly overestimates the value of $\Lambda$. The
$L^{-2/3}$ finite size correction (\ref{Airy}) derived in the harmonic
approximation improves the comparison, but quantitative agreement
is achieved only using the refined expression 
(\ref{eqnDeltaLambdaNew}), which is proportional to $L^{-1/2}$. 

\begin{figure}
\begin{center}
\includegraphics[scale=0.5]{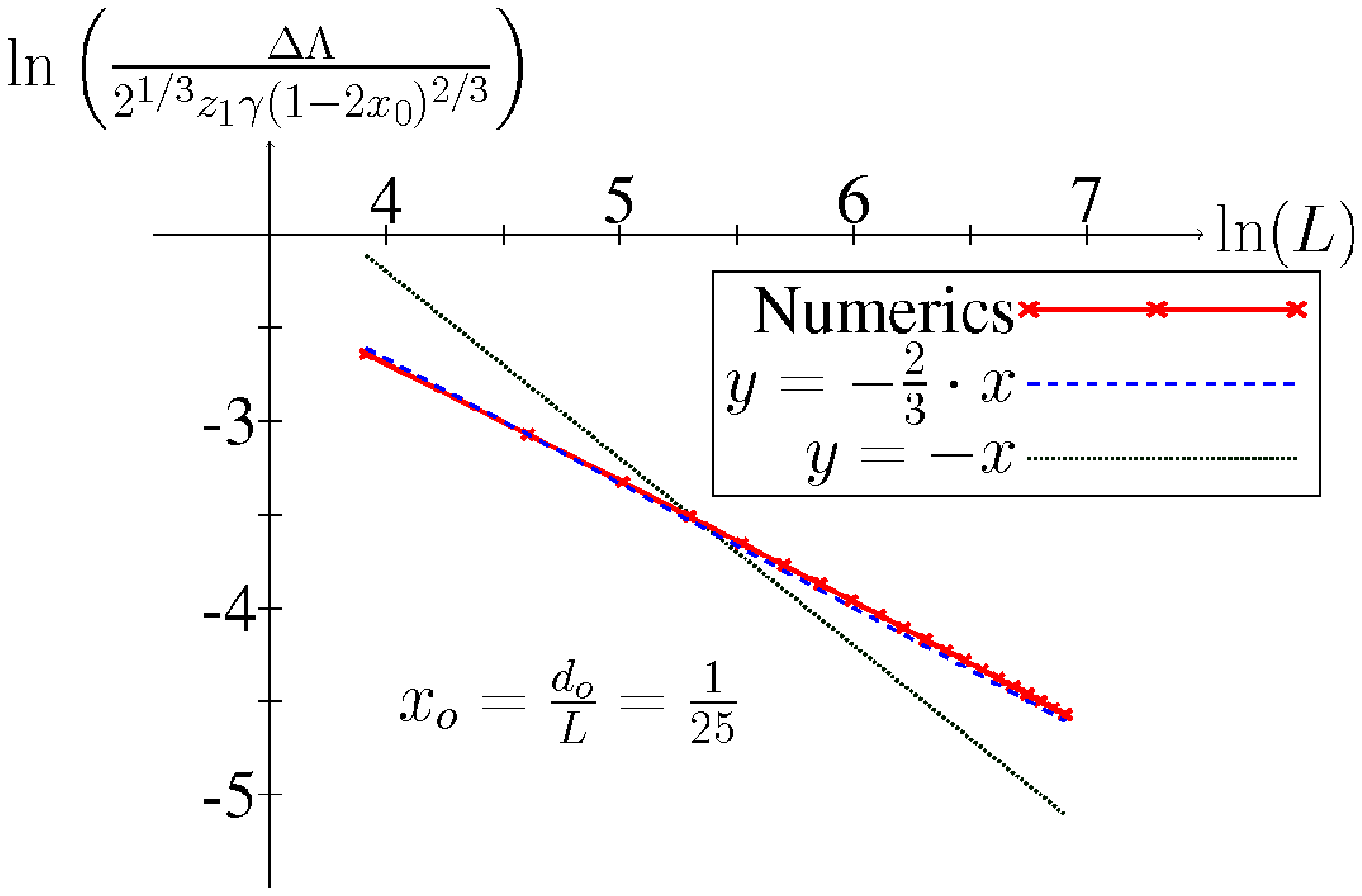}
\includegraphics[scale=0.5]{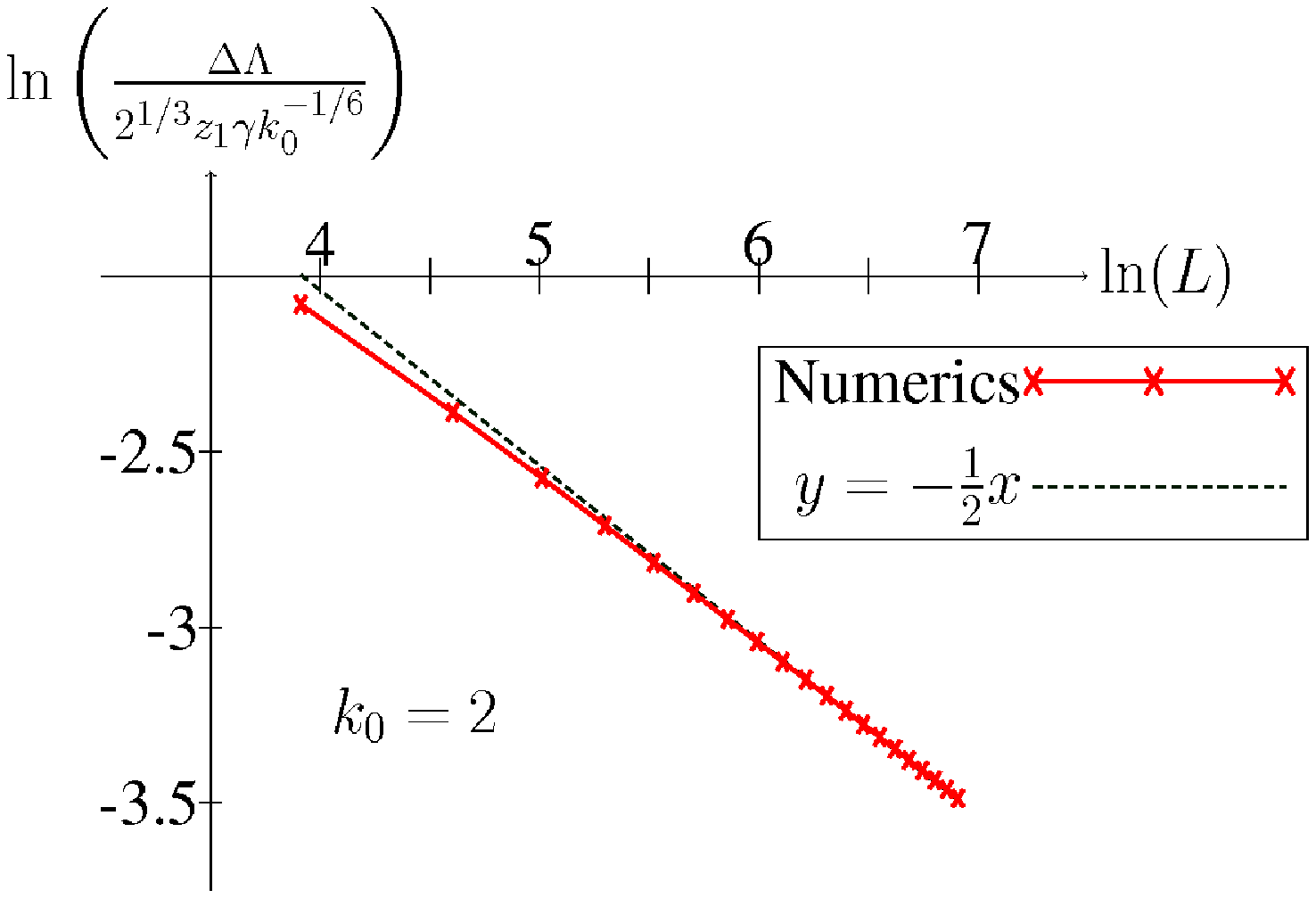}
\caption{Illustration of the power of the sequence length in the correction term for fixed relative and absolute plateau width. $\Delta\Lambda$ is the numerical value for the growth rate $\Lambda_{num}$ less the value obtained from the maximum principle, eq. (\ref{maximum}). For fixed relative plateau width, as well as for fixed absolute width, the numerics show the same exponent of the sequence length as the corresponding analytical result.}
\label{picexponent}
\end{center}
\end{figure}

Figure \ref{picwoc_von_k} shows a similar comparison for the 
critical plateau height $w_0^c$.  Here the
prediction (\ref{GHthresh}) of GH is also included and seen to match
the numerical outcome only poorly, whereas the MP result with the
finite size correction (\ref{eqnDeltaLambdaNew}) produces excellent
agreement. Finally, in left panel of figure \ref{picexponent} we verify that the 
finite size correction $\Delta \Lambda$ indeed varies as $L^{-2/3}$
when $L$ is increased at fixed \textit{relative} plateau width $x_0$. The right panel shows the corresponding $L^{-1/2}$ dependence for fixed absolute plateau width $k_0$.

\section{Fitness landscapes with competing peaks}
\label{Selection}

\begin{figure}
\begin{center}
\includegraphics[scale=0.8]{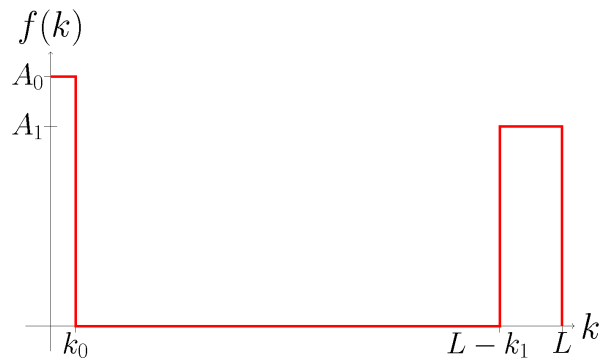}
\caption{Illustration of a fitness landscape with two plateaus. This type of landscape is used to investigate the influence of height and (relative) broadness of the plateaus on the population fitness $\Lambda$.}
\label{pictwoplateaus}
\end{center}
\end{figure}

\subsection{The selection transition} 

Since we have validated the analytical results of section \ref{Continuum} via numerical studies, we can now apply the analytical theory to the 
phenomenon of the survival of the flattest as explained in the introduction. To be specific, we want to find out
whether a broad plateau outcompetes a smaller but higher one even in the limit of long sequences.

In the literature this question
has already been discussed to some extent by Schuster and Swetina \cite{Schuster1988}.
The question can be answered by investigating a fitness landscape consisting of two fitness plateaus at the opposing ends of the Hamming space (see figure \ref{pictwoplateaus}). 
As was shown in \cite{Schuster1988}, for small $\mu$ the interference between the two
plateaus is negligible when they are separated by a few mutational
distances. The stationary state of the system is therefore to a very good approximation determined by the comparison between the population growth rates associated
with each of the two plateaus in isolation. 

Observing the center of mass of the population as function of the
mutation rate and for fixed sequence length, we find two types of
transitions. The first one is a jump of the population from the higher
to the broader plateau, which we will refer to as the
\textit{selection transition} \cite{Tarazona1992} taking place at a mutation rate $\mu_s$. 
The second transition is the well-known error threshold taking place at $\mu_{tr}$, where the population becomes uniformly spread in sequence space.
To analyze these transitions, an order parameter is needed. A convenient quantity is the population averaged ``magnetization'' defined by  
\begin{equation}
\label{M}
M= 1-2\langle x \rangle  \in [-1,1] \;\; \textrm{with} \;\;
\langle x \rangle = \frac{1}{L} \sum_{k=0}^L k P^\ast_k. 
\end{equation}
If the whole population consists only of master sequences, the magnetization is $M=1$. If only the inverse master sequence is present, the magnetization becomes $M=-1$. For a uniform distribution in sequence space 
(delocalized population) the magnetization is $M=0$. Thus we can distinguish the qualitatively different states of the population in the two plateau landscape by considering the population averaged magnetization $M$ as a function of $\mu$.

\begin{figure}
\begin{center}
\includegraphics[scale=0.45]{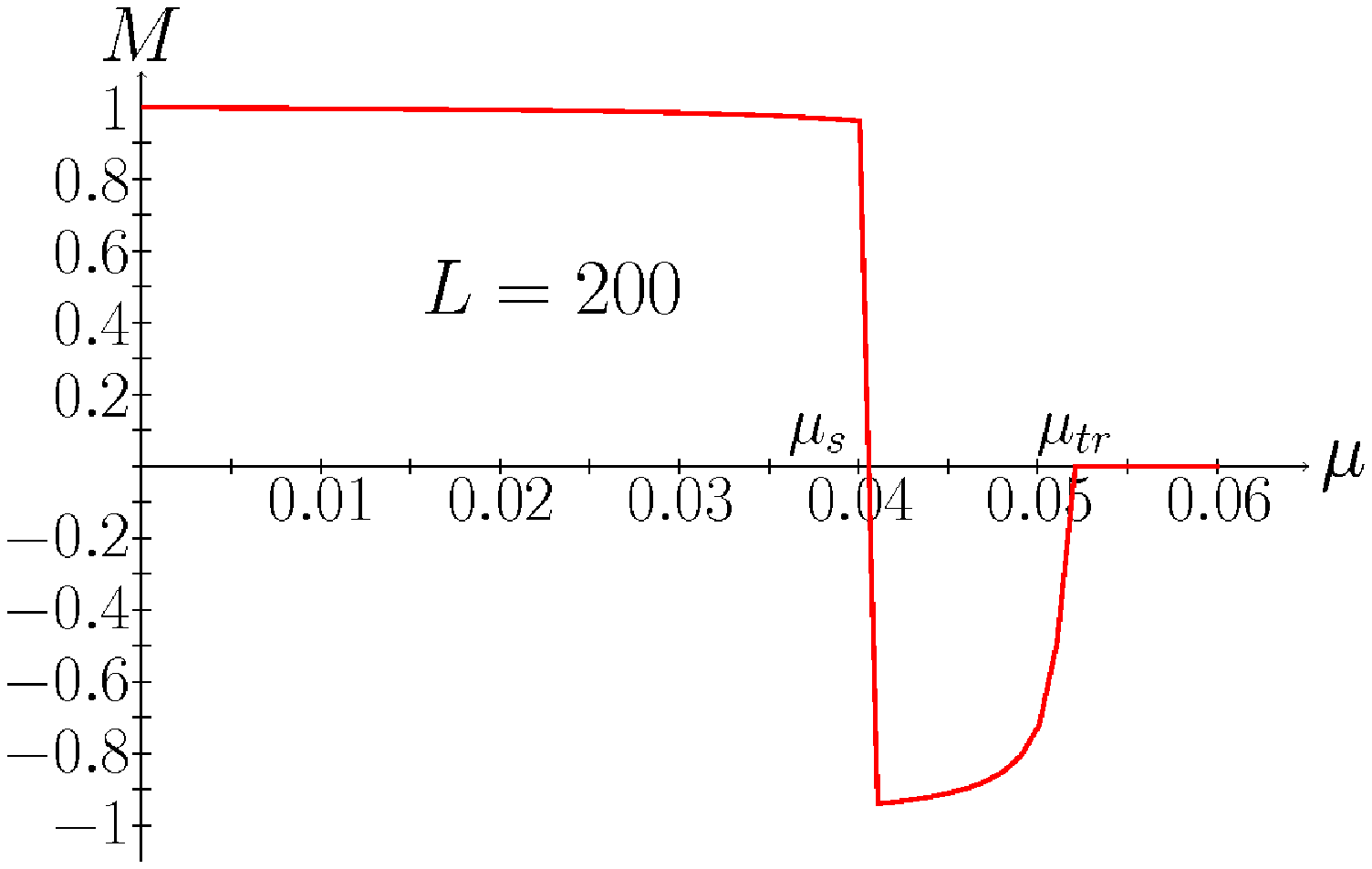}
\includegraphics[scale=0.45]{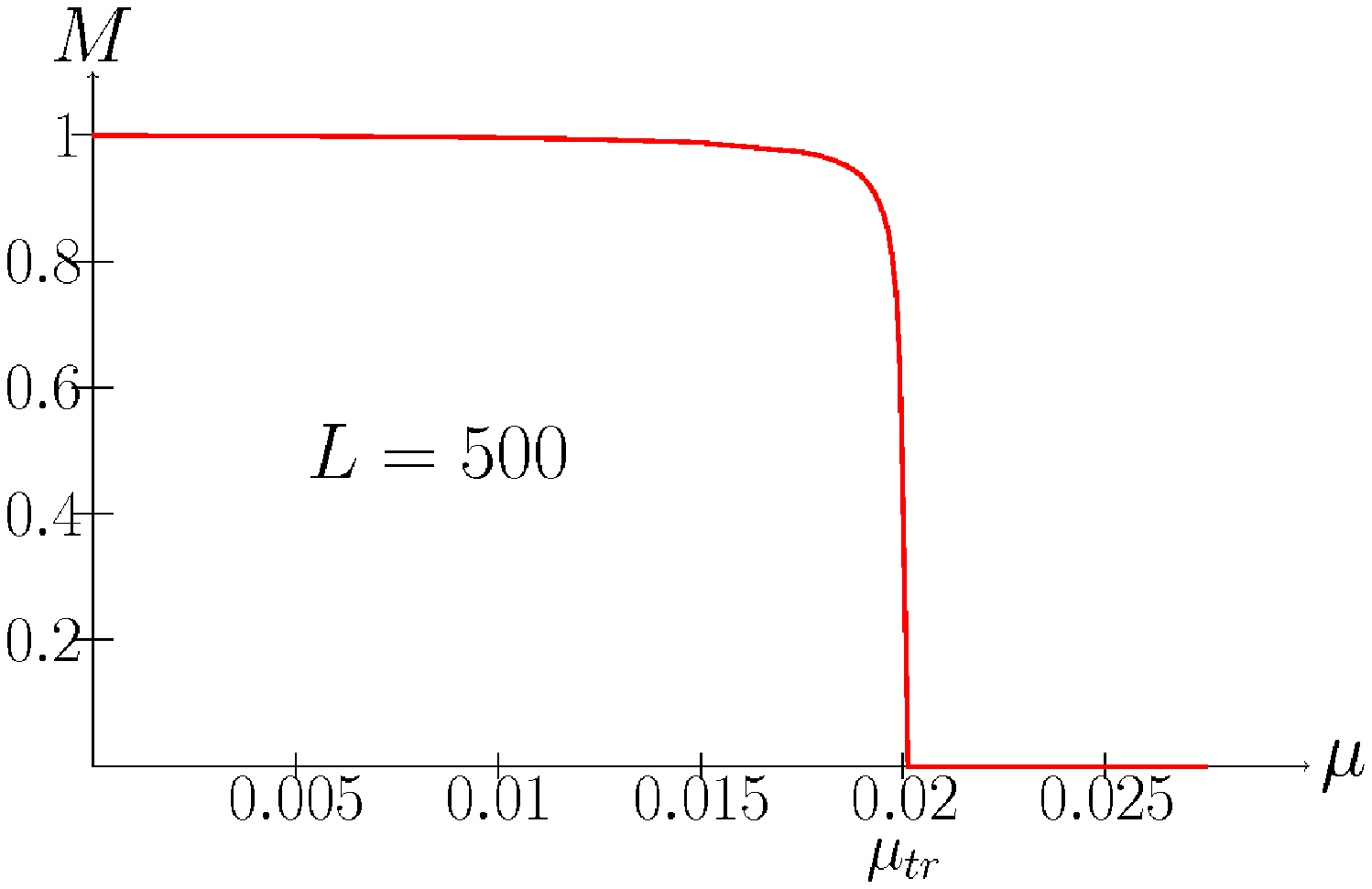}
\caption{Order parameter $M$ as function of the mutation rate $\mu$ per site for a fitness landscape with a high plateau at $k=0$ and a broad plateau at $k=L$. For short sequence lengths one can observe a hopping of the population from the higher to the broader plateau and then a delocalization (left picture). For long sequences, one only observes the delocalization transition from the higher fitness plateau (right picture). The hopping between the plateaus we call the \textit{selection transition}. It takes place at mutation rate $\mu_s$. The delocalization transition, also called error threshold, takes place at a mutation rate $\mu_{tr}$. The underlying fitness landscape is $w_k=10\cdot\Theta(1-k)+9\cdot\Theta(k-(L-2))$.}
\label{picmvonmuL200}
\end{center}
\end{figure}

As can be seen from figure \ref{picmvonmuL200}, the selection
transition (the jump between the two plateaus) is sharp even for
finite sequence lengths, whereas the error threshold is a continuous
transition for finite sequence length and only becomes sharp in the
limit of infinite sequence length \cite{Tarazona1992}. With growing sequence length the two critical mutation rates $\mu_s$ and $\mu_{tr}$ become smaller and also approach each other until, 
at a critical sequence length $L^\ast$, the selection transition completely disappears. For sequences longer than $L^\ast$, the population delocalizes directly from the high, narrow peak and the low,
broad plateau is never substantially populated.

\begin{figure}
\begin{center}
\includegraphics[scale=0.8]{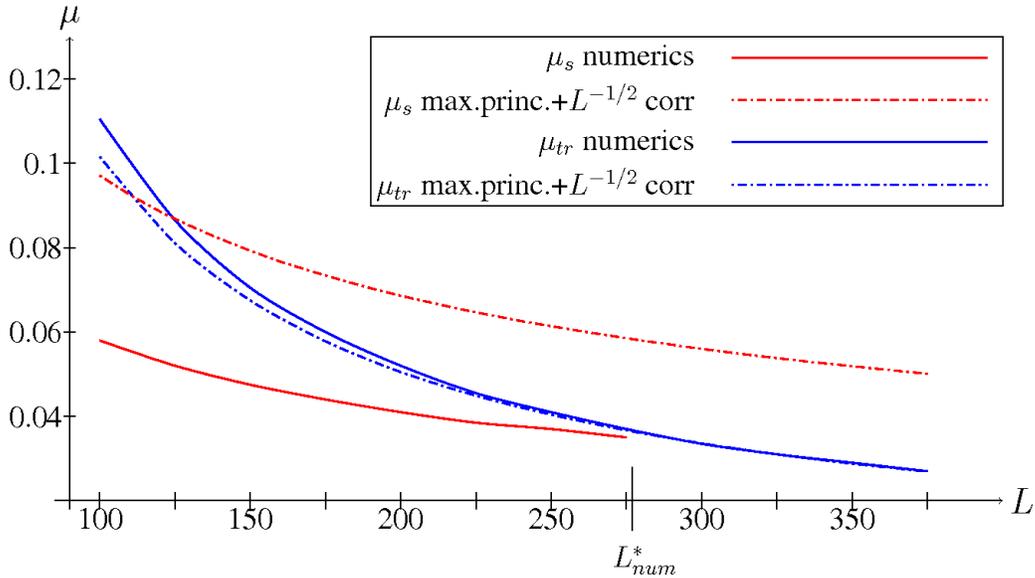}
\caption{Critical mutation rate $\mu_s$ of the selection transition and $\mu_{tr}$ of the error threshold for the fitness landscape $w_k=10\cdot\Theta(1-k)+9\cdot\Theta(k-(L-2))$ as functions of the sequence length. The two lines cross at a critical sequence length $L^{\ast}$. The selection transition is observed only for sequence lengths smaller than $L^{\ast}$. The numerical data are compared
to analytic predictions based on the MP including the $L^{-1/2}$-correction term. As before, the numerical values have been obtained by calculating the magnetization of the population and determining for each sequence length the mutation rate where the magnetization jumps.}
\label{picmusmutr}
\end{center}
\end{figure}

With the help of the maximum principle, this surprising behavior can be easily understood. 
Using (\ref{maximum}), the selection threshold is obtained by equating the population mean fitness for
the two competing peaks, which yields
\begin{equation}
\mu_s=\frac{w_0-w_1}{2\left(\sqrt{k_1L-k_1^2}-\sqrt{k_0L-k_0^2}\right)}
\approx \frac{w_0 - w_1}{2(\sqrt{k_1}-\sqrt{k_0})} L^{-1/2},
\label{mu_s}
\end{equation}
where $k_0,k_1 \ll L$ has been assumed in the last step.
On the other hand, the error threshold $\mu_{tr}^{(i)}$ associated
with plateau $i=0,1$ is determined by the vanishing of the corresponding
principal eigenvalue $\Lambda_i$, which gives
\begin{equation}
\mu_{tr}^{(i)} =\frac{w_i}{L\left(1-2\sqrt{k_iL-k_i^2}\right)} \approx
\frac{w_i}{1 - 2 \sqrt{k_i/L}} L^{-1}. 
\label{mu_tr}
\end{equation}
The different scaling of the two types of thresholds with sequence
length implies that for large $L$ the error threshold of the higher
peak is encountered before the selection threshold, which therefore
is no longer observable. The  
critical sequence length $L^{\ast}$ where the selection transition
vanishes can be estimated by equating the approximate expressions 
(\ref{mu_s}) and (\ref{mu_tr}), which yields
\begin{equation}
\label{Last}
L^\ast \approx \frac{4 (w_0 \sqrt{k_1} - w_1 \sqrt{k_0})^2}{(w_0
    - w_1)^2}.
\end{equation}
Following \cite{Schuster1988,Tarazona1992}, in our numerical work we
have considered short plateaus, $k_0 = 1$ and $k_1=2$,
with relative fitness values $w_1/w_0 = 0.9$, for which 
(\ref{Last}) give $L^\ast \approx 106$. Comparison with the numerical
values for the selection and error thresholds in figure \ref{picmusmutr}
shows that this significantly underestimates the value of $L^\ast$;
moreover, the agreement between theory and numerics is not
substantially improved by using the full expressions for the 
principal eigenvalues $\Lambda_0$ and $\Lambda_1$, including the
$L^{-1/2}$-correction derived in subsection \ref{secimprovement}.  
This is not surprising, as the continuum approach developed in section
\ref{Continuum} cannot be expected to be quantitatively accurate for plateaus
sizes of order unity. 

For completeness we mention that for plateau widths scaling with the
sequence length (such that $x_0 = k_0/L$ and $x_1 = k_1/L$ are kept
fixed as $L \to \infty$) the selection transition is maintained at a
fixed value of $\gamma$ 
\cite{Saakian2006}.

\subsection{The ancestral distribution}

In addition to the equilibrium population distribution $P^\ast_k$
attained at long times, we can also consider the \textit{ancestral
  distribution}, 
the equilibrium distribution of the backward time process, as
introduced by Baake and collaborators \cite{Hermisson2002,
  Baake2007}. The ancestral distribution $a_k$ gives information on the
origin of the equilibrium population and is obtained as the product of
the right eigenvector $P^\ast_k$ and the left eigenvector $P^{\ast
  \ast}_k$ of the mutation-selection matrix defined through
(\ref{Eigen}), $a_k \sim P^\ast_k \cdot P^{\ast \ast}_k$.

\begin{figure}
\begin{center}
\includegraphics[scale=0.8]{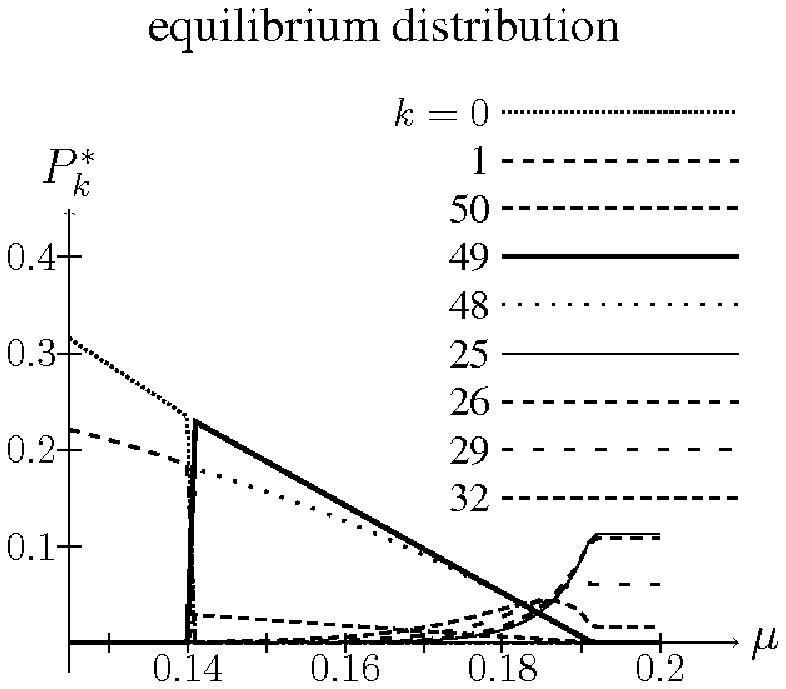}
\includegraphics[scale=0.75]{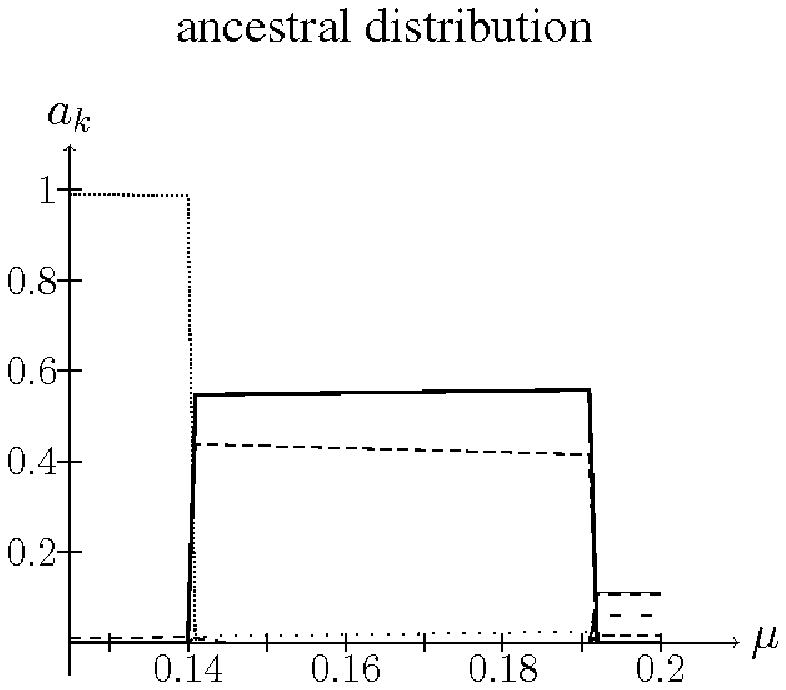}
\caption{The dominant entries of the equilibrium and ancestral
  population distribution as a function of the mutation rate per site,
  calculated numerically. The underlying fitness landscape is the same
  two-plateau landscape used in figures
\ref{picmvonmuL200} and \ref{picmusmutr}. The sequence length is chosen as $L=50$. Occupation fractions are plotted only for the most populated Hamming classes. The two distributions undergo phase transitions at the same mutation rates, but at the error threshold the ancestral distribution undergoes a discontinuous transition, while for the equilibrium distribution the transition is continuous.}
\label{picancestral}
\end{center}
\end{figure}

For the fitness landscape with two competing peaks, we find an
ancestral population that is either located on one of the plateaus or
uniformly distributed in sequence space
(figure \ref{picancestral}). The transitions beween these states are all
of first order. The continuous character of the error threshold
transition of the equilibrium population distribution, as opposed to
the discontinuous transition of the ancestral distribution, can be
explained by the growing mutational pressure affecting the population
on the plateau and driving it towards the middle of the Hamming
space. Mutations cause the population to "leak out" from the
plateau. Nevertheless, the individuals maintaining the population 
and compensating for the mutational loss are the ones with highest
fitness, which are located on the plateau and make up the ancestral distribution.

Before closing the analysis of plateau-shaped fitness landscapes, we
want to mention the connection between our description and the popular
language of Ising chains or semi-infinite Ising models \cite{Leuthaeusser1987,Baake2001}. In the
Ising picture, the ancestral distribution becomes the bulk
distribution on a semi-infinite two-dimensional (spatial or
spatio-temporal) lattice, and the equilibrium distribution becomes the
distribution in the surface layer. This analogy can be seen very
clearly in the paper by Tarazona \cite{Tarazona1992}, where the
different orders of the transitions in the two distributions are
explained in terms of surface wetting.

\section{Epistasis and the error threshold}
\label{epistasis}

So far, we have discussed robustness of phenotypes using plateau-shaped fitness landscapes, which are a special case of the class of epistatic fitness functions. We now want to discuss the latter in a more general framework. Epistasis describes the non-linear dependence of the fitness function on the number of mismatches $k$ \cite{Phillips2000}. Every additional mismatch is penalized harder 
(synergistic epistasis of deleterious mutations) or less hard (diminishing epistasis) than the previous one.
Here we address the effect of epistasis on the existence of an error threshold, which is defined for our purposes
as a singularity in the dependence of the population mean fitness on the mutation rate. In general (but not always, see 
below) such a singularity is associated with a discontinuous jump in the location of the most populated genotype. 

Following Wiehe \cite{Wiehe1997} we consider the class of permutation invariant (Malthusian) epistatic fitness functions
\begin{equation}
\label{wepi}
w_k=w_0-bk^{\alpha},
\end{equation}
where $k$ is again the Hamming distance to the master sequence and $b > 0$.
The epistasis exponent $\alpha$ takes the value $\alpha=1$ in the non-epistatic case, while 
$\alpha > 1$ and $\alpha < 1$ produces landscapes with synergistic and diminishing epistasis, respectively. 
For $\alpha \to 0$ (\ref{wepi}) reduces to the sharp peak landscape $w_k = w_0 - b(1 - \delta_{k,0})$.
It is well known that an error threshold exists for $\alpha\rightarrow0$, but not for $\alpha=1$ \cite{JK2006}. Neglecting backward mutations, Wiehe argued in \cite{Wiehe1997} that an error threshold emerges whenever $\alpha<1$. 
In the following we show that, based on the maximum principle (\ref{maximum}), the critical value of the epistasis exponent below which an error threshold develops is in fact $\alpha=1/2$. 

As before, we work in the scaling limit $L \to \infty$ and $\mu \to 0$ with the mutation rate per sequence 
$\gamma = \mu L = const$.
In order to cast (\ref{wepi}) into the form (\ref{fit}) required for the application of the maximum principle, we write 
\begin{equation}
w_k=f(x)=w_0-\tilde{b} x^{\alpha},\;\;\;\; \mathrm{with}\;\;x=k/L,\;\;\;\; \tilde{b}=bL^{\alpha}
\end{equation}
and the limit $L\rightarrow\infty$ should be combined with $b\rightarrow 0$, such that $\tilde{b}=const$. 
Since $b$ can be interpreted as a kind of selection coefficient, we are thus considering a situation where
both the mutation rate (per site) and the selection forces are small.  
Applying the maximum principle (\ref{maximum}) to this landscape, the mean fitness $\Lambda$ of the population in the equilibrium state is given by
\begin{equation}
\Lambda = \max_{x \in [0,1]} \{w_0-\tilde{b} x^{\alpha} - \gamma [1 - 2 \sqrt{x(1 - x)}]\} \equiv
\max_{x \in [0,1]} \lambda(x),
\label{lambdaepistatic}
\end{equation}
where $\lambda(x)$ is the function inside the curly brackets.

To find the condition under which the maximum is attained inside the interval $x\in (0,1)$, we set 
$d \lambda/dx = 0$, yielding the condition
\begin{equation}
\frac{\gamma}{\tilde{b}}(1-2x)= \alpha x^{\alpha-1/2}\sqrt{1-x}.
\end{equation}
For $\alpha>1/2$ the right hand side is a convex function which vanishes at $x=0,1$, with an infinite slope at $x=1$. As a consequence, there exists always a unique solution $x_c\in (0,1)$ for any value of $\gamma/\tilde{b}$, which describes the location
of the population for $L \to \infty$. The location varies smoothly from $x_c = 0$ for $\gamma/\tilde b \to 0$ to 
$x_c \to 1/2$ for $\gamma/\tilde b \to \infty$, and there is no error threshold. However, for $\alpha<1/2$ the right hand side diverges at $x=0$, and there is no solution for small $\gamma/\tilde{b}$. The function $\lambda(x)$ is then monotonically decreasing, which implies that the maximum in (\ref{lambdaepistatic}) is located at the boundary point $x=0$ over a finite interval of $\gamma/\tilde b$. 
Increasing $\gamma/\tilde b$ the function $\lambda(x)$ develops a local maximum, which eventually exceeds the boundary
value $\lambda(0) = w_0 - \gamma$. At this point the population discontinuously delocalizes to an interior point
$x_c \in (0,1)$. The error threshold condition is of the form
$\gamma/\tilde{b} = g(\alpha)$ where the function $g(\alpha)$ is not
explicitly available. This translates into the expression
\begin{equation}
\mu_{tr}=\frac{\gamma_{tr}}{L} = \frac{\tilde b g(\alpha)}{L} = g(\alpha) b L^{\alpha - 1},
\end{equation}
for the critical mutation rate $\mu_{tr}$. This scaling of $\mu_{\tr}$
with $L$ was also obtained in \cite{Wiehe1997}.
In the sharp peak limit $\alpha \to 0$ the threshold occurs at $\gamma/\tilde b =
\gamma/b = 1$, which implies that $g(0) = 1$.
On the other hand, for $\alpha = 1/2$ the expansion of $\lambda(x)$ near $x = 0$ reads
\begin{equation}
\label{xsmall}
\lambda(x) \approx w_0 - (\tilde b - 2 \gamma)x^{1/2} - \gamma x^{3/2},
\end{equation}
which shows that $g(1/2) = 1/2$. For $\gamma/\tilde b > 1/2$ an interior maximum appears 
at $x_c = (2 - \tilde b/\gamma)/3$, which moves \textit{continuously} away from $x=0$. 
In the language of phase transitions, $\alpha = 1/2$ can thus be viewed as a critical
endpoint terminating the line of discontinuous phase transitions that occur for
$\alpha < 1/2$.

\begin{figure}
\begin{center}
\includegraphics[scale=0.8]{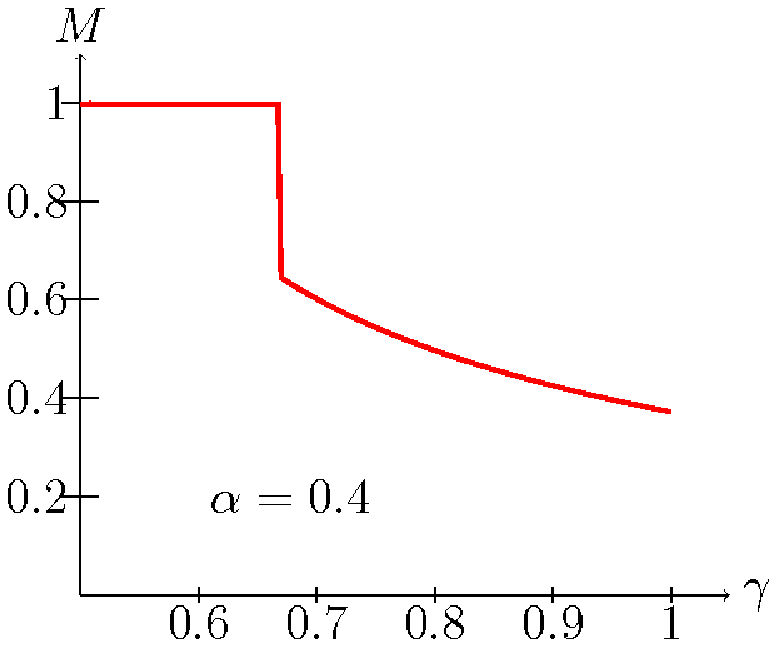}
\includegraphics[scale=0.8]{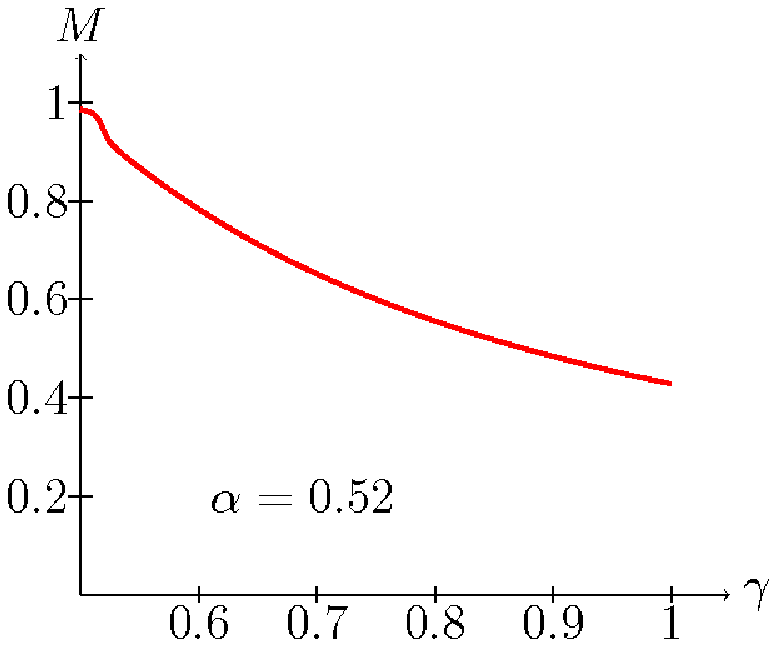}
\caption{Magnetization as a function of mutation rate for the fitness landscape (\ref{wepi}) with  epistasis exponent $\alpha=0.4$ and $\alpha=0.52$, respectively. For $\alpha=0.4$ the magnetization undergoes a discontinuous jump, whereas for $\alpha=0.52$, it changes smoothly. Calculations have been done for a sequence length of $L=500$.
}
\label{picepistaticmag}
\end{center}
\end{figure}

\begin{figure}
\begin{center}
\includegraphics[scale=0.8]{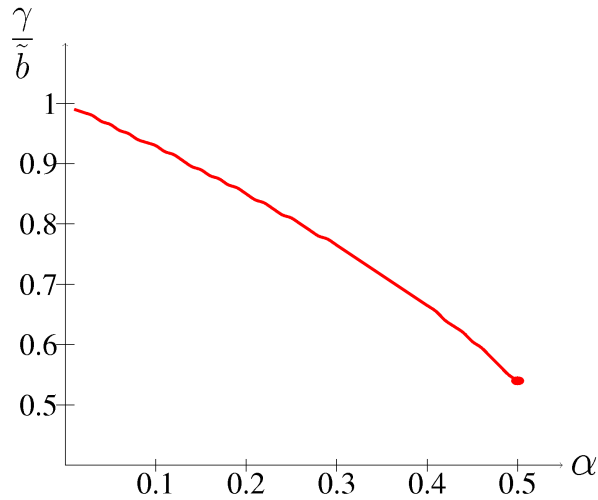}
\caption{Numerically determined phase diagram for the epistatic fitness landscape (\ref{wepi}). At the thick line the population undergoes a first order phase transition from a state localized at $x_c=0$ 
(below the line) to a delocalized state $x_c > 0$ (above the
line). This line terminates in a second order phase transition at
$\alpha=1/2$. The deviation from the prediction
$\gamma/\tilde{b}=g(1/2)=1/2$ at $\alpha=1/2$ is due to finite
sequence length corrections. For all larger values of the epistasis
exponent, $\alpha>1/2$, the population changes smoothly. Calculations
have been performed for a sequence lengths of $L=750$. The slight
modulation of the line is due to the finite numerical step size.
}
\label{picepistaticnum}
\end{center}
\end{figure}

These predictions are fully confirmed by numerical calculations for finite sequence length.
Figure \ref{picepistaticmag} illustrates
the existence of an error threshold for $\alpha<1/2$ and its absence for $\alpha>1/2$
by showing the behavior of the magnetization $M$ as a function of $\gamma$ for two 
different cases. The magnetization displays a non-analytic jump for $\alpha<1/2$ and 
varies smoothly for $\alpha>1/2$. In figure \ref{picepistaticnum} we
show the error threshold as a function $\gamma/\tilde b = g(\alpha)$,
which interpolates between the limits $g(0) = 1$ and $g(1/2) = 1/2$. 

It should have become clear that the special role of $\alpha = 1/2$ derives from the fact that for this value
the leading order behavior of the fitness function for small $x$ matches 
that of the ``entropic'' term $\sim \sqrt{x(1-x)}$ in the maximum principle (\ref{maximum}).  
Since a similar term appears also for general alphabet sizes [see (\ref{maximumA})], the considerations
of this section hold in that case as well.

\section{Conclusions}
\label{Conclusions}
In this paper we discussed the properties of epistatic fitness landscapes, with particular emphasis on mesa landscapes describing mutational robustness of phenotypes, which have been studied previously in the context of regulatory motifs \cite{Gerland2002,Berg2004}. As population evolution model we used the continuous time Crow-Kimura model, which is a quasispecies model for asexual and haploid organisms, and analysed its stationary states for sequences consisting of two letters. 
As explained in Appendix B, it is straightforward to generalize our results to sequence alphabets of general size.
Similarly, the extension to discrete time dynamics can be carried out by replacing the Malthusian fitness landscape
$w_k$ by its Wrightian counterpart $W_k \sim \exp(w_k)$ \cite{JK2006,Saakian2006,Park2006}.

We reviewed two existing approaches \cite{Gerland2002,Hermisson2002,Saakian2006} to this problem and 
explained the discrepancy between their predictions by extending 
the approach of Gerland and Hwa \cite{Gerland2002} beyond the harmonic approximation.
Based on a quantum mechanical analogy we derived a novel finite size correction term to the maximum principle
of \cite{Hermisson2002}, which 
significantly improves the agreement with numerical calculations.
Our central result is that 
the relative number of tolerable mismatches $x_0=k_0/L$ is the relevant parameter for the fitness effect
of mutational robustness, and we provide accurate formulae for its quantitative evaluation. 
As a consequence, we showed that the selection transition first described by Schuster and Swetina \cite{Schuster1988}
disappears for long sequences. 
 
Finally, in section \ref{epistasis}, we discussed more general forms of epistatic fitness landscapes with regard to
the existence of an error threshold. Based on the results of \cite{Hermisson2002,Baake2007} we improved on earlier work 
\cite{Wiehe1997} and showed that diminishing epistasis [$\alpha < 1$ in the fitness function (\ref{wepi})] is 
not a sufficient condition for an error threshold to occur. 

\section*{Acknowledgements}

We are grateful to Alexander Altland, Ellen Baake, Ulrich Gerland, Kavita Jain,
Luca Peliti and David Saakian for useful discussions. This work was supported by 
DFG within the Bonn Cologne Graduate School of
  Physics and Astronomy,  SFB 680 \textit{Molecular basis of
  evolutionary innovations} and SFB/TR12 \textit{Symmetries and
  Universality in Mesoscopic Systems}.

\section*{Appendix A: The large deviations approach}

We start by symmetrizing the eigenvalue problem (\ref{Eigen}). The
discrete analogue of the transformation (\ref{psi}) is 
\begin{equation}
\label{Qk}
Q_k = {L \choose k}^{1/2} P^\ast_k,
\end{equation}
which leads to 
\begin{equation}
\label{QkEigen} \hspace*{-1.5cm}
\Lambda Q_k = (w_k - \gamma) Q_k + \mu \sqrt{(L-k)(k+1)} \, Q_{k+1}
+ \mu \sqrt{(L-k+1)k} \, Q_{k-1}.
\end{equation}
Following \cite{Saakian2007}, we now perform the continuum limit by 
making a large deviations ansatz for $Q_k$, 
\begin{equation}
\label{Qlarge}
Q_k = Q_{xL} = \psi(x) = \exp[-\epsilon^{-1} u(x)]
\end{equation}
with $\epsilon = 1/L$. Inserting this into (\ref{QkEigen}) one finds
\begin{equation}
\label{HJ}
(\Lambda - f(x) + \gamma) \psi = 2 \gamma \sqrt{x(1-x)} \cosh[u']
\psi.
\end{equation}
Cancelling $\psi$ on both sides yields a Hamilton-Jacobi equation for the
`action' $u(x)$, with $u' = du/dx$ playing the role of a canonical
momentum \cite{Saakian2007}. In order to cast (\ref{HJ}) into the form
of a Schr\"odinger equation, we expand the momentum-dependent factor
to quadratic order, $\cosh(u') \approx 1 + (u')^2/2$, 
and make use of the relation
\begin{equation}
\label{u'}
(u')^2 = \epsilon^2 \psi^{-1} \frac{d^2 \psi}{dx^2},
\end{equation}
which follows from (\ref{Qlarge}) to leading order in $\epsilon$.
Inserting this into (\ref{HJ}) results in (\ref{schroedinger}).

\section*{Appendix B: General alphabet size}

Here we show how our results
generalize to the case where the symbols in the genetic sequence
are taken from an alphabet of $A > 2$ letters (for nucleotide
sequences $A = 4$). We assume a uniform 
point mutation rate $\mu$ connecting any two of the $A$ possible states of a site in 
the sequence, and 
a fitness function $w_k$ that depends on the relative number of mismatches
according to (\ref{fit}). It is then straightforward
to see that the basic eigenvalue problem (\ref{Eigen}) generalizes to 
\begin{eqnarray}
\label{EigenA}
(\Lambda - w_k) P^\ast_k = \\  
= (A-1) \gamma \left[ \frac{k+1}{A-1} P^\ast_{k+1} + 
(L - k + 1) P^\ast_{k-1} - \left(L - k + \frac{k}{A-1}\right)
 P^\ast_k \right]. \nonumber
\end{eqnarray}
Applying the results of
\cite{Baake2005,Garske2004a,Garske2004b}
to this problem we find that, asymptotically for large 
$L$, the principal eigenvalue is given by the maximum principle
\begin{equation}
\label{maximumA}
\Lambda = \max_{x \in [0,1]} \left\{f(x) - (A-1)\gamma \left[ \left(1-
    \frac{(A-2)x}{A-1} \right) - \frac{2 \sqrt{x(1 - x)}}{\sqrt{A-1}}
\right] \right\}.
\end{equation}
For the case of the mesa landscape (\ref{mesa}) this implies that 
the population is localized near the optimal genotype for 
$w_0 > w_0^c$ with 
\begin{equation}
\label{w0cA}
w_0^c = \gamma (A - 1) \left[1 - \frac{(A-2)x_0}{A-1} - \frac{2
\sqrt{x_0 (1-x_0)}}{\sqrt{A-1}} \right].
\end{equation}
The right hand side is a monotonically decreasing function of $x_0$
which vanishes at $x_0 = 1 - 1/A$. For fixed $x_0$ it is an increasing
function of $A$.

\section*{References}
\bibliographystyle{unsrt}
\bibliography{robust.bib}

\end{document}